\documentclass[fleqn,usenatbib]{mnras}

\usepackage{newtxtext,newtxmath}
\usepackage[T1]{fontenc}
\usepackage[utf8]{inputenc}

\DeclareRobustCommand{\VAN}[3]{#2}
\let\VANthebibliography\thebibliography
\def\thebibliography{\DeclareRobustCommand{\VAN}[3]{##3}\VANthebibliography}

\usepackage{graphicx}	
\usepackage{amsmath}	
\usepackage{siunitx}
\usepackage{booktabs}

\newcommand{\ebv}{{$E$(B~$-$~V)}}
\newcommand{\ebtvt}{{$E$(B$_T - $V$_T$)}}

\newcommand{\bp}{B$_P - $R$_P$}

\newcommand{\btvt}{B$_T - $V$_T$}

\newcommand{\rdnote}[1]{{\textcolor{red} }}

\providecommand{\parallax}{\ensuremath{\varpi}}

\providecommand{\sigparallax}{\ensuremath{\sigma_{\varpi}}}
\providecommand{\fpu}{\ensuremath{\parallax/\sigparallax}}

\title[Be stars in Gaia DR3]{A Photometric Comparison of B and Be stars using Gaia DR3}

\author[I. C. Radley et al.]{Isaac C. Radley$^{1}$\thanks{E-mail: isaac.radley@gmail.com},
Ren\'{e} D. Oudmaijer$^{2,1}$,
Miguel Vioque$^{3}$,
and Jonathan M. Dodd$^{1}$ 
\\
$^{1}$School of Physics \& Astronomy, University of Leeds, Woodhouse Lane, Leeds LS2 9JT, UK\\
$^{2}$ Royal Observatory of Belgium, Ringlaan 3, 1180 Brussels, Belgium \\
$^{3}$European Southern Observatory, Karl-Schwarzschild-Str. 2, 85748 Garching bei München, Germany\\
}

\date{Accepted 2025 April 14. Received 2025 April 14; in original form 2024 August 28}

\pubyear{2024}

\begin{document}

    \label{firstpage}
    \pagerange{\pageref{firstpage}--\pageref{lastpage}}
    \maketitle

    \begin{abstract}
        Previous studies have observed significant photometric differences between non-emission B-type and classical Be stars, however the precise mechanism responsible for these differences is unclear. This study combines the Bright Star Catalogue with Tycho and Gaia photometry to create a homogeneous sample of 1015 of the closest and brightest B and Be-type field stars with 90 per cent of objects at distances $<\,$500\,pc. 
        Due to their proximity, the extinction towards these objects is very low, ensuring we minimise any obfuscation in the reddening correction and final photometry. We present our findings in both Tycho and Gaia photometry through colour magnitude diagrams and present intrinsic colours and absolute magnitudes for each spectral type. We find Be stars are on average $\sim\,$0.5 magnitudes brighter in both Gaia $G$ and Tycho V$_T$ compared to non-emission B stars of the same spectral type. Additionally, we find tentative evidence that Be stars are redder in Gaia \bp, particularly for the earlier types, but have similar Tycho \btvt \ colours. 
        We test the effects of gravitational darkening due to rapid rotation and binarity on the photometry of our sample and find both to be insufficient to explain the observed photometric differences between B and Be stars.  We conclude that the most likely mechanism responsible for the observed photometric differences is the combined effect of the circumstellar disc and stellar evolution up the Main Sequence, with the disc dominating early-types and evolution dominating late type stars.
    \end{abstract}

    \begin{keywords}
        stars: emission-line, Be -- stars: circumstellar matter -- stars: fundamental parameters -- Hertzprung-Russell and colour-magnitude diagrams.
        
    \end{keywords}

    \section{Introduction} \label{sec:Intro}

        Emission B stars (Be) have often been seen as an enigma within the stellar physics community due to their unusual photometric properties and disc structures. The current working definition of a Be star is a non-supergiant early-type star which currently or previously exhibited Balmer lines in emission \citep{Jones2022,Rivinius_review, BeSS}. In addition to Balmer emission, Be stars are regularly observed with a significant infrared excess which is typically thought to originate from free-free, bound-free and recombination emission within a gaseous, circumstellar disc \citep{Dougherty1994,2Mass_IntrinsicColours, IRExcessStudy}. The discs are small, of order milli-arcseconds, and were  first imaged interferometrically in the nineties of the last century \citep{Quirrenbach1997}. 

        The circumstellar disc is one of the key features of Be stars giving rise to many of the properties which make Be stars so unique.
        The mechanism which forms the gaseous disc cannot be described singularly, however, rapid rotation of the central star must be a significant factor \citep{GravityDarkening}. Be stars are one of, if not, the most rapidly rotating class of non-degenerate stars \citep{Zorec2016}. However, the origin of their fast rotation is still a matter of debate. Either the stars were born rapidly rotating, spun-up during their  evolution  or they gained mass and angular momentum during an interaction with a companion star   (e.g. \citealt{Rivinius_review}). The latter hypothesis has gained traction over the last years, with evidence emerging that Be stars are often accompanied by stripped stars that have lost their envelopes (e.g. \citealt{Naze2022,ElBadry2022, Klement2024,Klement2025}) whereas  comparative studies into B and Be star binarity appear to confirm this from a statistical point of view \citep{Dodd2024}. 

In addition, Be stars are known to exhibit photometric and spectroscopic variability \citep{Kelt_Variability} on timescales as small as 10 minutes \citep{BeBinaries} up to months and decades \citep{BeSS}. Previous studies have found short to mid term variations in the Hipparcos H$_P$ band \citep{Turon1992}, were typically less than 0.3 magnitudes \citep[see e.g.][]{Hubert1998}.
Of particular note is the fact that circumstellar discs have been found to build-up and then disperse on timescales of months to years (e.g. \citealt{Okazaki2007,Haubois2012,Jones2013}). Therefore, the unique observational properties which distinguish Be stars are their Balmer emission lines, strong infrared excess and photometric variability.
        
For many decades it has been known that Be stars may be as bright or brighter than their non-emission counterparts \citep[see e.g.][]{Merrill33,Jung70}. Studies such as \citet{Zorec91} and \citet{BriiotAbsV} further confirm the enhanced brightness of Be stars but also find differing relations between the brightness enhancement and spectral type/temperature. Additionally, studies by \citet{Weg2V,Zhang} and \citet{Rot_SpType} which relied on Hipparcos \citep{Hipparcos} measurements confirmed that Be stars appear to be intrinsically brighter than regular B stars, appearing to sit 0.5-1\,mag above the main sequence. Further afield, studies of Be stars in clusters in the local group (e.g. \citealt{Milone2018,Dufton2022}) or beyond (\citealt{Schootemeijer2022});  have also confirmed that Be stars are often redder than their non-H$\alpha$ emitting counterparts. 

Several explanations for these photometric differences have appeared in the literature. Among these we find that gravity darkening of rapidly rotating B stars can lead to substantially brighter objects when seen pole-on and redder when seen edge-on (e.g. \citealt{GravityDarkening}). On the other hand, the contribution of the circumstellar disc to the total photometry has been mentioned as well. The exact photometric impact of the disc is dependent on various parameters such as stellar rotation and inclination \citep[see e.g.][]{Poeckert1978,Moujtahid1999}. Finally, the Be phenomenon could occur when B stars evolve up the Main Sequence (\citealt{Zorec2005}) and a possible spin-up of the star can then give rise to the formation of a disc
(e.g. \citealt{Mombarg2024}). When evolving up the Main Sequence, the stars can become brighter and redder. We will return to these hypotheses in more depth later in this paper. 

With the advent of Gaia \citep{Gaia2016} and its ever increasing parallax precision (30 per cent increase between last two data releases, \citealt{GaiaDR3}) it is timely to revisit the photometric properties of Be stars. Here we use the highest precision parallaxes we have available from Gaia Data Release 3 (DR3; \citealt{GaiaDR3}) to accurately determine distances and therefore their intrinsic observational characteristics, enabling us to accurately place Be stars in the colour magnitude diagram (CMD). Therefore, any uncertainties in \ebv \ and $A_V$ should be kept to a minimum.

The best way of minimising the impact of any extinction correction is to have as low an extinction towards the objects as possible. As Be stars are not found to be surrounded by dust, this naturally implies we need to study the closest examples that suffer the least reddening by interstellar dust. To this end, we will use the Bright Star Catalogue and its Supplement (BSC, \citealt{BSC,BSC_Supplement}) containing the brightest and nearest B stars. These catalogues are magnitude limited, and typically complete for $V$ $\approx$ 7. For the stars under consideration, this translates into typical distances of 100s of pc.   Moreover, we need to avoid the use of spectral type based extinction corrections given that Be stars may appear redder by almost 2 spectral subclasses, leading to an overestimate of their extinction corrected absolute magnitudes.  This can  be minimised by applying different, non-spectral type based, corrections such as extinction maps, which will be most accurate for the nearest, least reddened objects. 

In this paper we restrict our studies exclusively to B spectral type (B and Be) stars of luminosity class V. We note that late O and early A type stars can be considered to be very similar to, if not, completely fulfil the Be definition \citep{BeSS}, while non-supergiant Be stars such as sub-giants and giants (luminosity classes IV and III respectively) are sometimes considered as well. These restrictions help to reduce any misclassifications or uncertainties in given spectral types \citep{Rivinius_review}.

This paper is organised as follows. In Section \ref{sec:Method}, we discuss the sample selection and extinction treatment. Section \ref{sec:Results} describes the photometric properties of B and Be stars and presents colour-magnitude diagrams for both {\it B$_T$,V$_T$} magnitudes (in the Tycho system) as well as for the Gaia ({\it G, B$_P$, R$_P$}) passbands. We discuss the results and possible mechanisms responsible for the differing photometric properties of B and Be stars in Section \ref{sec:Discussion}, while we conclude in Section \ref{sec:ConcludingRemarks}.

\section{Sample preparation} \label{sec:Method}
\subsection{Sample Selection}\label{sec:SampSelect}
We combine the Bright Star Catalogue 5th Revised Ed. from \citet{BSC} with its supplementary material \citep{BSC_Supplement} leading to a master catalogue containing 11,721 sources which is complete down to $V$=7.10. Within this catalogue spectral types are given for the majority of sources (99.9  per cent) with 2,424 categorised under the B-type spectral class regardless of luminosity, sub-type and emission classifications. Henceforth when referring to the BSC we will only consider B and Be spectral types. The complete BSC (original plus Supplement) was cross matched within 5" against Gaia Data Release 3 (DR3) \citep{GaiaDR3} using the 'Basic Query' functionality on the Gaia archive.

454 of these BSC sources had more than one DR3 match, with the majority of these BSC sources having 2 possible Gaia counterparts. As we are using the BSC, every source is, by definition, very bright, and with the exception of two stars, the brightest match was selected.  HD~36862 and HD~147934 were selected by hand due to the proximity of an even brighter nearby star leading to a catalogue of 2396 B stars. 

\begin{figure}
    \centering
    \includegraphics[width=\linewidth]{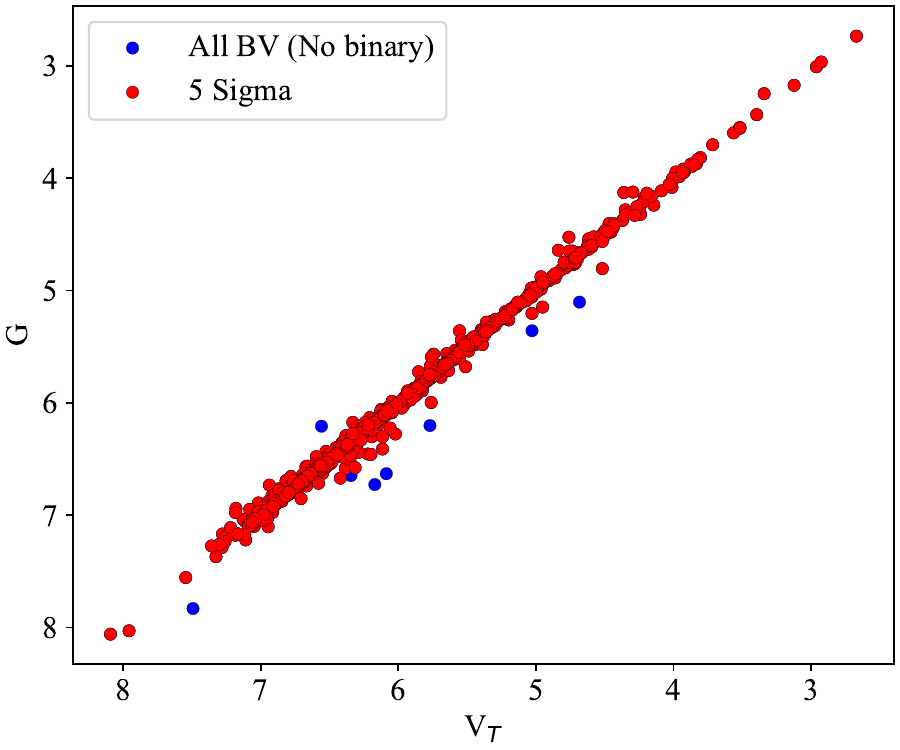}
    \caption{Comparison of the apparent V$_T$ band magnitude against the apparent G band magnitude for the non-binary and final (5$\sigma$) samples in blue and red, respectively. 
    \label{fig:G_Vt_Comp_Clean}}
\end{figure} 

In order to obtain modern, homogeneous photometry of the sample, we use the Tycho-2 catalogue (and supplementary catalogues, \citealt{Hog2000}). Tycho uses a bespoke $B_T, V_T$ photometric system, where the filters bear close resemblance to the Johnson $B,V$ bands. Our sample of B stars was cross-matched with the Tycho catalogue using a 1'' search radius. We found 21 sources had 1 duplicate and from which the brightest of the two was selected. We extended the search radius from 1'' to 5'' for  HD~79469, HD~41534, HD~181869, HD~108767, HD~11503 and HD~222661 leading to a sample of 2389 B stars.                    

Continuing  with only the BV and BVe stars we removed objects where the spectral type signified a binary star (e.g. when a '+' is included in the spectral type) as this would clearly affect their location in the colour-magnitude diagram. Additionally, as we require accurate distance estimates, we also require the most accurate parallaxes available. We therefore, implement the parallax cut, \fpu \ $>$ 5, as discussed in \citet{Bailer-Jones2021}
within our analysis, only considering sources with parallax measurements above this threshold. This led to a BV and BVe catalogue of 1027 stars, however not all objects had complete photometry in both Tycho and Gaia.  

 In order to ensure that photometric variability between the Tycho and Gaia observing epochs does not seriously affect the placement of the objects in the CMDs, we finalise our sample by only choosing sources which showed close agreement between $G$ and $V_T$ band magnitudes. We define this through computing $G-V_T$ and removing any sources beyond 5 standard deviations ($\sim 0.3$ mag) of the mean, leading to a final catalogue of 897 BV and 118 BVe stars. These objects are shown in Figure \ref{fig:G_Vt_Comp_Clean} where a very strong linear trend can be seen indicating that $G$ and $V_T$ magnitudes are similar for B-type stars. It may be interesting to point out that the 5$\sigma$ cut  resulted in the removal of only 12 objects, about 1\% of the total, while the 1$\sigma$ spread in the difference in magnitudes is small ($\sim$0.06 mag).

\begin{table}
    \centering
\begin{tabular}{|l|r|r|r|}
\hline
  \multicolumn{1}{|c|}{Spectral type} &
  \multicolumn{1}{c|}{B} &
  \multicolumn{1}{c|}{Be} &
  \multicolumn{1}{c|}{Be/B (\%)} \\
\hline
  B0V+B0.5V & 13 & 3 & 23.1\\
  B1V+B1.5V & 48 & 10 & 20.8\\
  B2V+B2.5V & 105 & 29 & 27.6\\
  B3V & 97 & 19 & 19.6\\
  B4V & 32 & 11 & 34.4\\
  B5V+B5.5V & 111 & 13 & 11.7\\
  B6V & 40 & 9 & 22.5\\
  B7V & 46 & 4 & 8.7\\
  B8V+B8.5V & 147 & 11 & 7.5\\
  B9V+B9.5V & 258 & 9 & 3.5\\
  Total & 897 & 118 & 13.2\\
\hline \end{tabular}
\caption{BV and BVe populations for each spectral type of our final sample.}
\label{tab:BDwarf_pop}
\end{table}
Finally, in Table \ref{tab:BDwarf_pop} we show the number of B and Be sources as function of their spectral type, as well as the B to Be fraction. We find that the Be fraction appears to decrease towards later spectral types (B7-B9). If we consider the total dwarf (V) population we find a Be fraction of 13.2 per cent which is slightly below the 15-20 per cent quoted in \citet{Rivinius_review}. The difference in the derived Be fraction could be caused by the transient nature of the Be phenomenon itself or the fact that our sample includes a large fraction of late-type B9 stars, where the occurrence rate is much lower (see Table~\ref{tab:BDwarf_pop}). 

\subsection{Extinction}\label{sec:Int_Extinction}

One of the main aims in this study is to minimise the effects of extinction towards each object in order to accurately determine their photometric properties. Our first step in minimising extinction was already made by constraining ourselves to the brightest and nearest objects. Normally, we can determine the extinction to stars by using intrinsic colours for their respective spectral type to determine the colour excess and thus, extinction. However, we wish to avoid any biases in the spectral classification of the Be stars that may be the result of their fast rotation and emission line spectrum. Therefore, we aim to estimate the extinction to the Be stars independently of their spectral type. One method to determine the extinction independently of stellar properties is to use a dust map which takes into account a star's position and distance. The dust map we have opted to use in this study is Stilism \citep{Stilism_Dust, Lallement18}. 

\begin{figure}
    \centering
    \includegraphics[width=\linewidth]{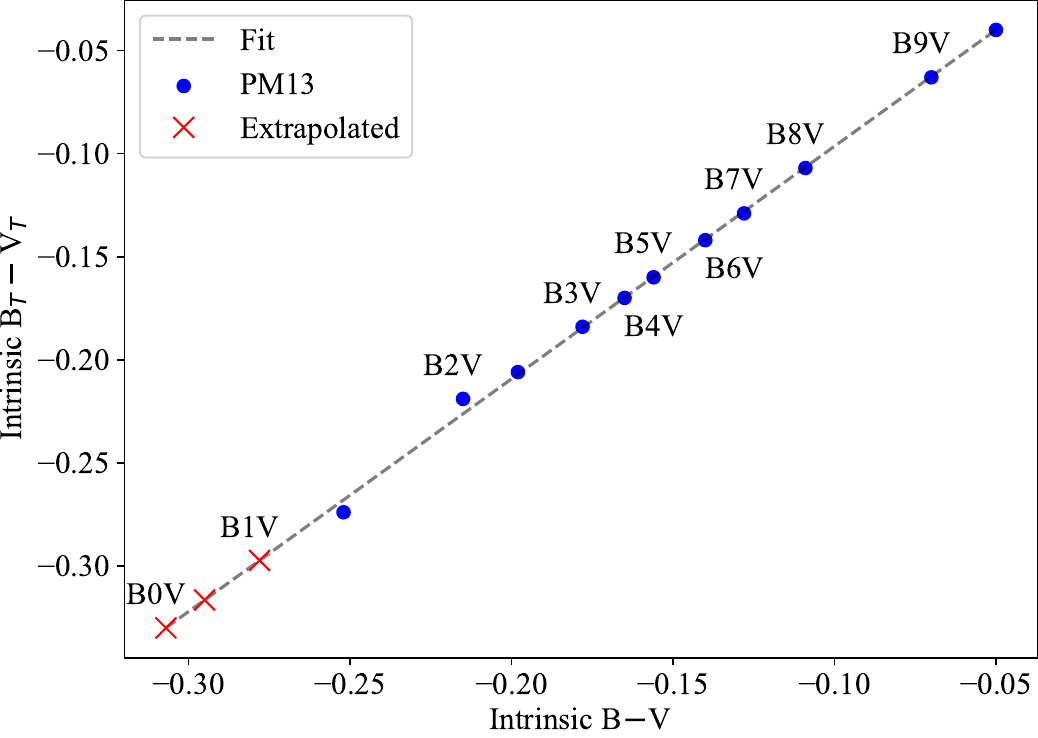}
    \caption{Intrinsic colours from \citet{Intrinsic_Colours_V} (PM13) shown as blue dots with B0-B1 extrapolated from the relationship between {\it B$_T-$V$_T$} and {\it B$-$V} shown as red crosses. We annotate the associated integer spectral types for each data point.}
    \label{fig:Johnson_tycho_extrapolator}
\end{figure}

In order to check whether this method returns valid line-of-sight extinction values, we  first derive the extinction for the non-emission B dwarf stars using the intrinsic \btvt \ values based on their spectral type taken from \citet{Intrinsic_Colours_V}. This table does not list intrinsic \btvt \ colours for the B0V, B0.5V and B1V spectral classes and we estimate these by comparison with  intrinsic $B-V$ colours.  When plotting the intrinsic \btvt \ against  $B-V$ as listed in \citet{Intrinsic_Colours_V} in Figure~\ref{fig:Johnson_tycho_extrapolator},
we find a gradient of 1.13~$\pm$~0.02 which indicates the similarity between the two photometric bands. We compute {B$_T-$V$_T$} values by extrapolating our best-fit line to find B$_T-$V$_T$ $\sim$ -0.33, -0.32 and -0.30 for B0, B0.5 and B1 respectively.

\begin{figure}
    \centering
    \includegraphics[width=\linewidth]{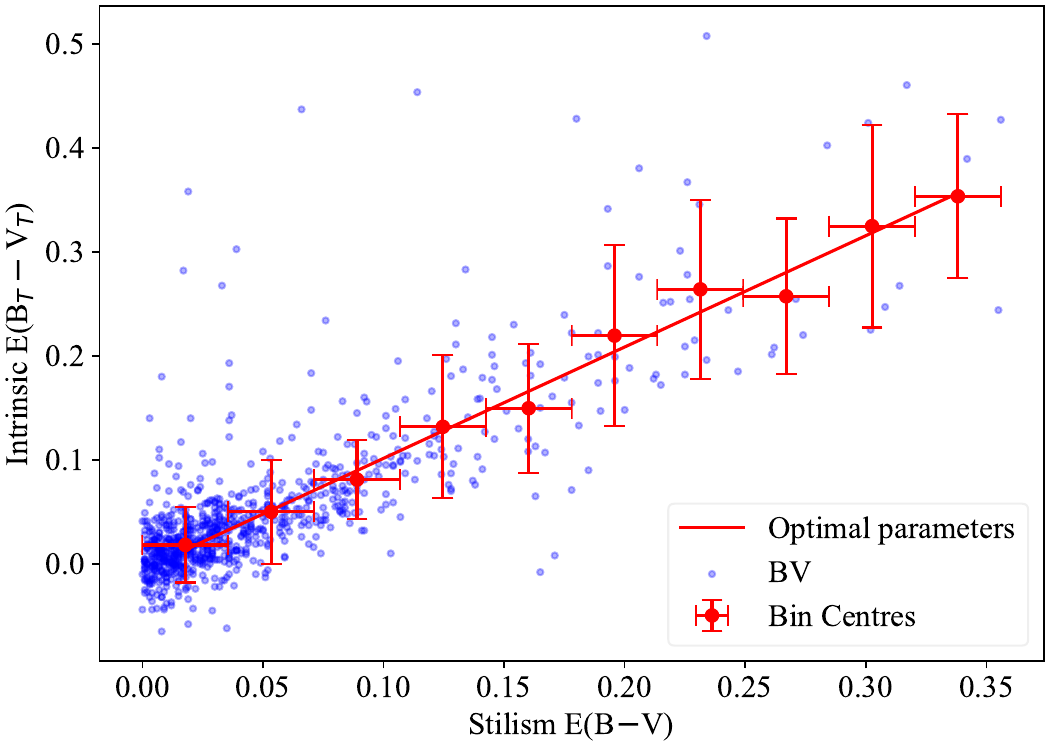}
    \caption{Non-emission BV stars (blue circles) with \ebtvt \ calculated based on the intrinsic colours for their respective spectral type compared to the interstellar extinction as tabulated by Stilism (see Section~\ref{sec:Int_Extinction}). We bin the data into 10 \ebv \ bins with centres indicated as red circles and represent the bin width and standard deviation within each bin as the X- and Y-errors respectively. We also show the best fit line in solid red.
    } 
    \label{fig:Stil_v_Theor_EBtVt_Comparison}
\end{figure}

We can now compare the extinction towards our B stars, computed using their intrinsic colour, with the extinctions estimated by Stilism. Figure~\ref{fig:Stil_v_Theor_EBtVt_Comparison} shows the \ebtvt \ computed using the spectral types against the (Johnson) \ebv \ listed by Stilism.  A strong correlation between the theoretical and interstellar extinctions is present indicating that Stilism is performing well in comparison to the intrinsic extinction calculations. All but two objects (not shown) have \ebv$_{\rm stilism}$ \ $<$ 0.4, which reinforces our initial assumption that the majority of BSC objects do not suffer from high extinction values. In the following analysis we will remove the two objects with Stilism \ebv \ $\sim$ 0.85 and \ebv \ $\sim$ 0.65, thus limiting the range in colour excesses to a maximum of 0.4 magnitudes.

\begin{figure*}
    \centering
    \includegraphics[width=0.45\linewidth]{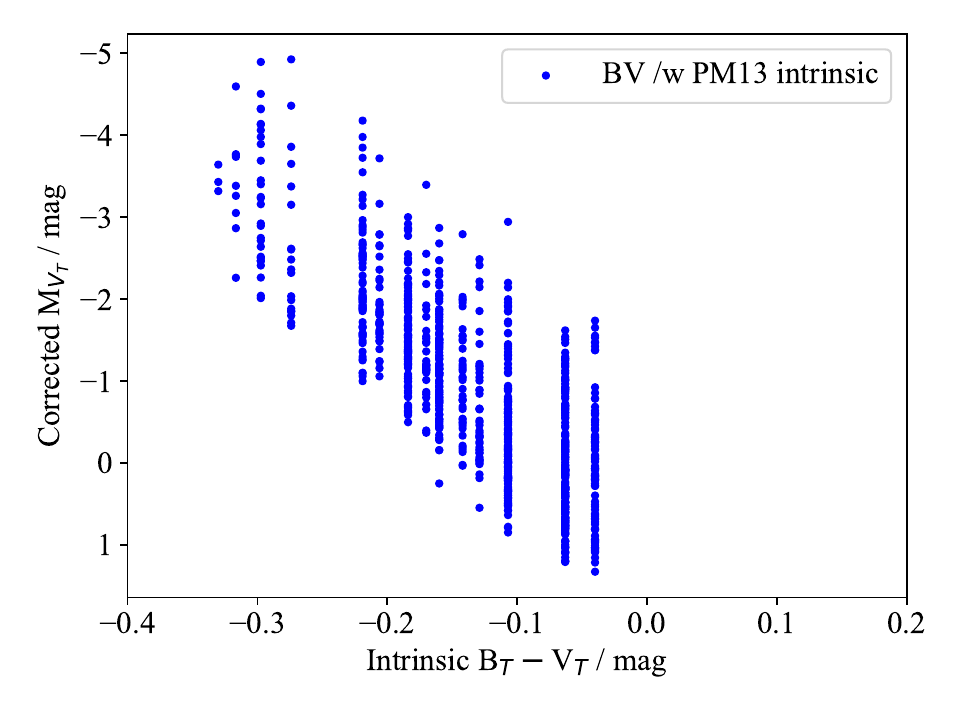}
    \includegraphics[width=0.45\linewidth]{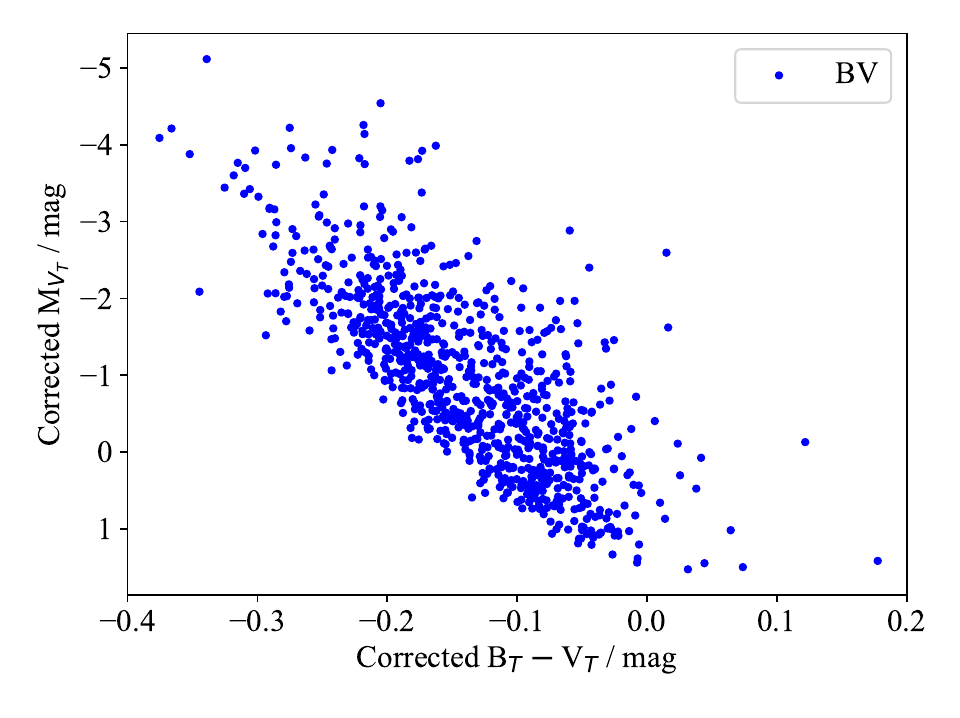}
    \caption{\textbf{\textit{Left:}} CMD of the B Main Sequence stars (BV) within the BSC \citep{BSC} shown as blue circles. The \btvt \ and $V_T$ magnitude values have been calculated using intrinsic colours from \citet{Mamajek02} and parallax distances from DR3 \citep{GaiaDR3}. Sources have been corrected using the intrinsic colours for their spectral type (see text) and have parallax/error $>$ 5. \textbf{\textit{Right:}} CMD of the B-stars (blue circles), now with the extinction corrected using the Stilism dust map \citep{Stilism_Dust,Lallement18} (see Section~\ref{sec:Int_Extinction}). Using the average parallax error of the BV objects, we find an average uncertainty of $\sim$0.02 mag in M$_{V_T}$}
    \label{fig:BV_Theor_stil_HRD}
\end{figure*}

In order to determine the relationship between the two extinction methods and hence obtain an estimation for uncertainties within Stilism, we divide the  Stilism \ebv \ range into 10 bins, and compute the mean and standard deviation of the corresponding intrinsic \ebv . Figure \ref{fig:Stil_v_Theor_EBtVt_Comparison} shows a clear linear correlation  along with the best fitting line which has a gradient of 1.07~$\pm$~0.04 and an intercept of -0.006~$\pm$~0.009. The dispersion on the individual extinction values is of order 0.04-0.1 magnitude, which may reflect the fact that stars of the same spectral type will have been assigned exactly the same intrinsic $B-V$ values, whereas in practice there must be a range of intrinsic colours associated with each spectral type. In fact, the steps in intrinsic colour typically vary around 0.02-0.03 magnitudes (see Figure~\ref{fig:Johnson_tycho_extrapolator}) which can account for a large amount of the scatter in Figure~\ref{fig:Stil_v_Theor_EBtVt_Comparison}. In the following we correct all objects for extinction using the Stilism provided \ebv \ multiplied by a correction factor of 1.07 as our estimate of \ebtvt.

Figure~\ref{fig:BV_Theor_stil_HRD} shows the Colour-Magnitude diagram for BV stars corrected for extinction using the intrinsic colours in the left panel\footnote{The absolute $V_T$-band is corrected for extinction using \\ $ A_{V_T} = R_{V_T} \times E(B_T-V_T)$, with $R_{V_T} \approx 3.44$ using the passband conversions given in \citet{Casagrande2014} and \citet{Casagrande2018}}, and using the Stilism corrected values in the right-hand panel. 
We find that in both cases the objects occupy similar absolute $V_T$-band and \btvt \ ranges, but for the Stilism corrected photometry the \btvt \ distribution is not limited to discrete values. Although some objects such as those blueward and redward of the Main Sequence are most likely outliers, it can be concluded that  Stilism performs well for the vast majority of objects.  This  validates applying the method to the Be stars without relying on their spectral types, that can be subject to bias, for the extinction corrections later in this paper.

\subsection{Sample Comparison}
\subsubsection{Properties}
We can now compare both B and Be samples to estimate any photometric biases that we may have introduced. The most obvious, in terms of observational impact, is the typical distance to each object. We find that on average our Be stars are found at a distance of 390~$\pm$~20~pc which is about 100 pc further away than our average B star distance of 272~$\pm$~6~pc. Since Be stars are considered to be rarer, we would also therefore expect them to be found at further distances than their non-emission counterparts. Additionally, although the difference in distance between the two samples is small it can have a drastic effect on the observed photometry, particularly if the object is in a high extinction environment. 

Therefore, in order to ensure that distance does not significantly impact our observations, we also compare the average \ebtvt \ colour excesses derived for both B and Be stars. Unsurprisingly, Be stars have a marginally higher \ebtvt \ with an average of 0.084~$\pm$~0.007~mag compared to the average B colour excess of 0.056~$\pm$~0.002~mag which is likely due to their larger distances. Once again, the differences between the two groups regarding colour excess is minimal and thus we do not expect significant observational biases from comparing the two groups.

\begin{figure*}
    \centering
    \includegraphics[width=0.45\linewidth]{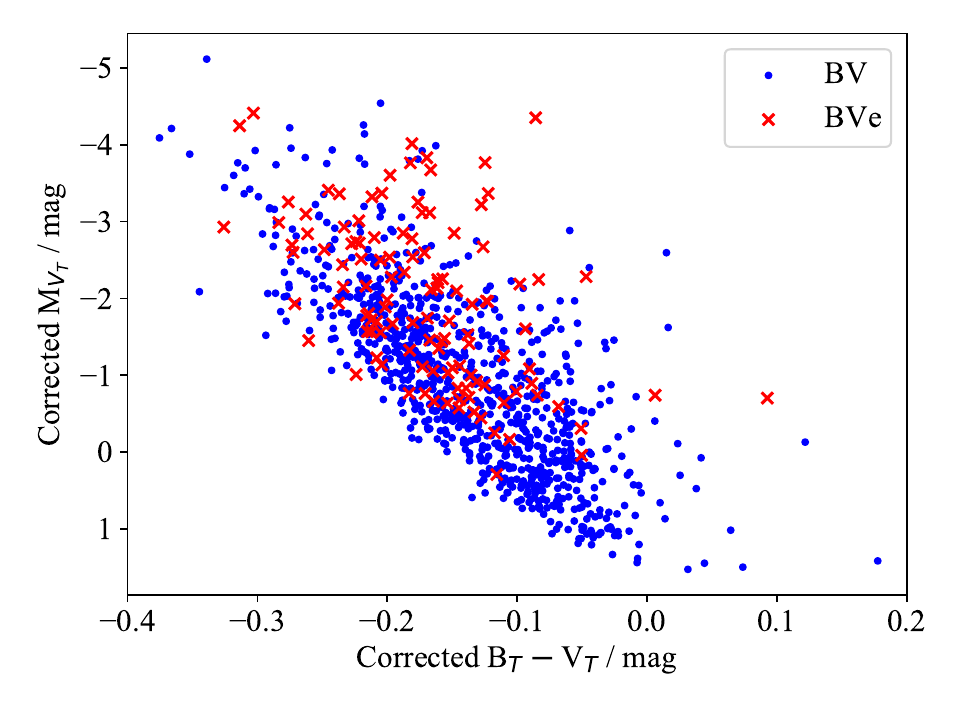}
    \includegraphics[width=0.45\linewidth]{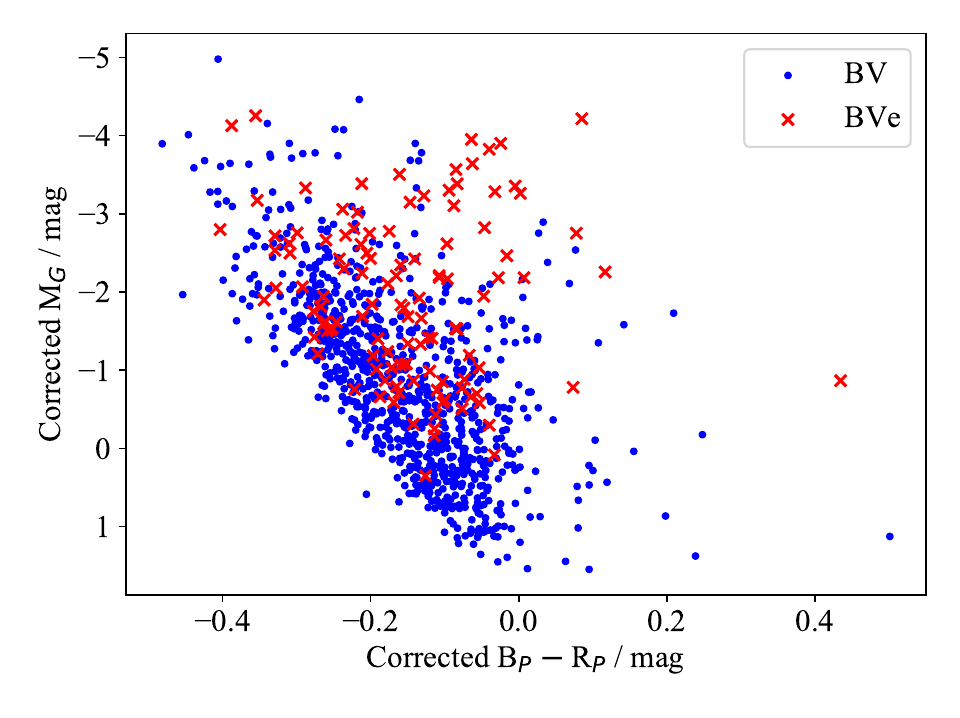}
    \caption{\textit{\textbf{Left:}} CMD of BV (filled circles) and BVe stars (crosses) in the BSC \citep{BSC}. The \btvt \ and $V_T$ magnitude values have been taken from Tycho \citep{Hog2000} and parallax distances from DR3 \citep{GaiaDR3}. Sources have been corrected using the Stilism dust map \citep{Stilism_Dust,Lallement18} (see text for details) and have parallax/error $>$ 5. \textbf{\textit{Right:}} The same but now for sources with extinction corrected B$_P -$ R$_P$, $G$. Using the average parallax error, we find an average uncertainty of $\sim$0.02 mag in M$_{V_T}$ and M$_{G}$ for both BV and BVe. Note that the Be stars are generally brighter and redder than the B stars. The latter is especially apparent for the Gaia colours.}
    \label{fig:StilHRD_B_BE}
\end{figure*}

\section{Results} \label{sec:Results}

Using the extinctions provided by Stilism, multiplied by 1.07 to match the extinction values retrieved for B stars, we now also correct the magnitudes and colours of the Be stars. Similarly, Gaia $G$ magnitudes and $B_P-R_P$ colours were corrected accordingly\footnote{Reddening-corrected Gaia photometric values are obtained by
converting the \ebtvt \ using the passband conversions given in \citet{ZhangExtinction} and \citet{Casagrande2018}. The total extinction in the Gaia $G$ band can be written as 
$    A_G \approx 2.488 \times E(B_T-V_T)$. 
The colour excess in $B_P-R_P$ is computed from $   E(B_P-R_P) \approx 1.327 \times E(B_T-V_T)$.
}. As before, we do not include sources with extinctions beyond \ebv$_{\rm Stilism}$~$>$~0.4. 

Figure \ref{fig:StilHRD_B_BE} shows the colour-magnitude diagrams for the Tycho and Gaia photometry respectively. In general we find that Be stars appear to be brighter both within a given spectral type and as a whole.  It would appear that no Be stars are present for M$_{V_T}$~$>$~0.3, which could be related to the very low number of B9Ve stars, the faintest B stars. 

Finally, we summarise the average Tycho photometry, M$_{V_T}$, \btvt \ , and Gaia photometry, M$_G$ and \bp \ , for each spectral subclass in Tables \ref{tab:BV_Avg_photometry} (B) and \ref{tab:BVe_Avg_photometry} (Be). In the following we compare our derived values to those found in the literature and investigate any differences between the B and Be stars in more detail.

\subsection{Photometric properties of B and Be stars}

\begin{table*}
    \centering
    \begin{tabular}{lcccccc}
\toprule
 & \multicolumn{3}{c}{B} & \multicolumn{3}{c}{Be} \\
 \cmidrule(lr){2-4}
 \cmidrule(lr){5-7}
Spectral Type & M$_{V_T}$ & W06 M$_{V}$ & SK81  M$_{V}$ &  M$_{V_T}$ & Z06 M$_{V}$ & W06 M$_{V}$ \\
\midrule

       B0V & -4.13 $\pm$ 0.83 &   -3.34 $\pm$ 0.33 & -4.00 & -3.62 $\pm$ 0.40 & --- & -3.81 $\pm$ 0.57 \\
       B1V & -3.06 $\pm$ 0.16 &   -2.95 $\pm$ 0.14 & -3.30 & -3.31 $\pm$ 0.17 & -3.23 $\pm$ 0.75 & -2.89 $\pm$ 0.28 \\
       B2V & -2.18 $\pm$ 0.09 &   -2.64 $\pm$ 0.09 & -2.50 & -2.67 $\pm$ 0.16 & -2.31 $\pm$ 0.22 & -2.64 $\pm$ 0.19 \\
     B2.5V & -1.93 $\pm$ 0.10 &    --- &  --- & -2.12 $\pm$ 0.21 & -2.53 $\pm$ 0.25 &   --- \\
       B3V & -1.59 $\pm$ 0.06 &   -1.61 $\pm$ 0.08 & -1.70 & -2.12 $\pm$ 0.18 & -2.01 $\pm$ 0.17 & -2.04 $\pm$ 0.24 \\
       B4V & -1.38 $\pm$ 0.11 &   -1.22 $\pm$ 0.12 &   --- & -1.91 $\pm$ 0.25 & -2.29 $\pm$ 0.21 & -1.93 $\pm$ 0.28 \\
       B5V & -1.18 $\pm$ 0.06 &   -1.15 $\pm$ 0.09 & -0.80 & -1.67 $\pm$ 0.26 & -1.34 $\pm$ 0.18 & -1.85 $\pm$ 0.42 \\
       B6V & -0.99 $\pm$ 0.11 &   -0.81 $\pm$ 0.10 & -0.50 & -1.18 $\pm$ 0.26 & -1.37 $\pm$ 0.28 & -1.40 $\pm$ 0.27 \\
       B7V & -0.64 $\pm$ 0.11 &   -0.63 $\pm$ 0.12 & -0.20 & -1.56 $\pm$ 0.37 & -2.06 $\pm$ 0.73 & -1.55 $\pm$ 0.76 \\
       B8V & -0.36 $\pm$ 0.07 &   -0.51 $\pm$ 0.06 &  0.10 & -0.87 $\pm$ 0.18 & -1.25 $\pm$ 0.21 & -0.86 $\pm$ 0.59 \\
       B9V &  0.02 $\pm$ 0.06 &    0.21 $\pm$ 0.04 &  0.50 & -0.84 $\pm$ 0.16 & -0.69 $\pm$ 0.42 & -0.86 $\pm$ 0.35 \\
     B9.5V &  0.01 $\pm$ 0.08 &    0.29 $\pm$ 0.08 &  --- & -0.35 $\pm$ 0.39 &  --- &  ---  \\

\end{tabular}
    \caption{Mean corrected M$_{V_T}$ values within each BSC assigned spectral type alongside their statistical error on the mean. We compare our B results with \citet{Weg2V} (W06) and \citet{MS_ZAMS} (SK81). Uncertainties are consistently calculated for both this work and W06 using the statistical error on the mean ($\frac{\sigma}{\sqrt{N}}$), where $\sigma$ is the dispersion. We compare the Be values to W06 and \citet{Zhang} (Z06), once again calculating errors as the statistical error on the mean.}
    \label{tab:LitComp}
\end{table*}

\subsubsection{Absolute magnitudes of the B stars} \label{sec:B_phot_Comp}

In the above, we determined extinction-corrected absolute {\it V$_T$} and Gaia $G$ band magnitudes for the most nearby B and Be dwarf stars for which we also have Gaia DR3 parallax based distances available. Due to our increased accuracy in the distance measurement, it may be interesting to see how these new values compare to earlier determinations of absolute magnitudes for B stars. Although there are  differences between the Johnson  and Tycho photometric systems, these are minimal.  Indeed, the very similar intrinsic colours in these systems, as shown in Figure \ref{fig:Johnson_tycho_extrapolator}, enable us to discuss the quantitative similarities between our results and those from the literature.

Table \ref{tab:LitComp} shows the average M$_{V_T}$ per spectral type compared with literature M$_V$ values. \citet{Weg2V} use Hipparcos parallaxes for their distances, while \citet{MS_ZAMS} is a classical resource listing fundamental parameters of stars. We find our results to be consistent with the literature and in particular with \citet{Weg2V}. The majority (73 per cent) of our results are consistent within $\sim 2 \sigma$ with the largest departure ($\sim 4 \sigma$) being found for B2.

\subsubsection{Absolute Magnitudes of Be stars}\label{sec:Be_phot_Comp}

Following the same sample selection criteria as for the B stars, and again restricting our analysis to luminosity class V Be stars, we can compare our derived photometric properties to other works in the literature. 

Similarly to the B stars, we find excellent agreement with \citet{Weg2V}'s values with the majority (6/10) of our average values being within less than 0.1 mag difference. \citet{Weg2V} includes both luminosity class IV and V within their average whereas we have restricted our results to just luminosity class V which may explain some of the differences between our results. We find a much greater difference with the results of \citet{Zhang} where 60 per cent of the spectral types have a discrepancy of 0.3 mag or greater. However, all of our results are still consistent within $\sim 1 \sigma$. We also note that \citealt{Zhang}'s M$_V$ values will on average be slightly overestimated due to their extinction correction method. The differences between our results and the literature will be further increased due to the difference in photometric bands although this should only be a minor effect \citep[see e.g.][]{Bessell2000}.

\raggedbottom

Finally, we note that this work also considers the photometric colours in \btvt \ and \bp \ for B and Be stars. However, as neither of these are reported in the literature, we do not compare our results here.

 \subsection{Photometric Comparison}\label{sec:B_Be_comp}

\begin{figure}
    \centering
    \includegraphics[width=\linewidth]{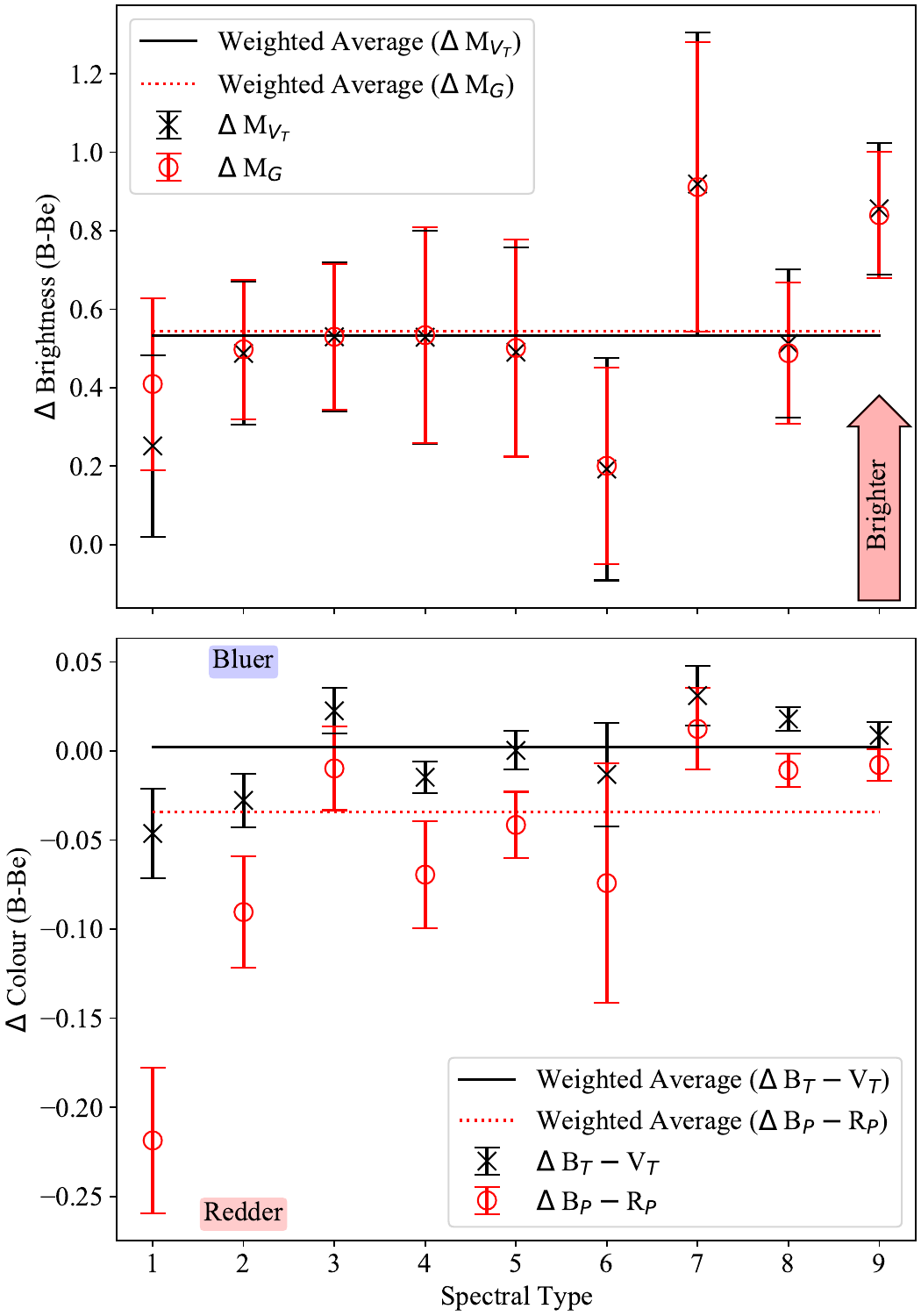}
    \caption{\textbf{\textit{Top:}} Difference in M$_{V_T}$ (Black crosses) and M$_G$ (Red open circles) for B and Be in each integer spectral type where errors have been propagated using the uncertainty on the average ${V_T}$ and $G$ magnitudes. The black solid and red dotted lines show the weighted average for $\Delta$M$_{V_T}$ and $\Delta$M$_{G}$ respectively. We also indicate the direction of increasing brightness difference of Be stars in the bottom right. \textbf{\textit{Bottom:}} Difference in \btvt \ (Black crosses) and \bp \ (Red open circles) for B and Be in each spectral type following the same error propagation as for the top panel. The black solid and red dotted lines show the weighted average for $\Delta$(\btvt) and $\Delta$(\bp) \ respectively. We also indicate the regions in which Be stars are considered redder or bluer than their non-Be counterparts.}
    \label{fig:PhotometryDifferences}
\end{figure}

In the following analyses we do not include half (e.g. B1.5V) spectral types, nor do we include B0 due to the small sample size of B0e stars (see Table \ref{tab:BVe_Avg_photometry}). If we consider the average properties of both populations we can quantify how bright Be stars appear compared to B stars. The top panel of Figure \ref{fig:PhotometryDifferences} shows the difference between our calculated average M$_{V_T}$ for B and Be stars. We find across all spectral types that Be stars are brighter in the visual with a minimum difference of 0.19 mag and a maximum increase of 0.92 mag. The majority of our results lie within the canonical range of brightness excess between 0.5-1 mag \citep{BriiotAbsV,Weg1V,Weg2V}. If we take a weighted average we find an average increase of 0.53 mag ($\sim7\sigma$) in the $V_T$ band for Be stars compared to their non-Be counterparts. 

We can additionally consider the absolute brightness difference in the Gaia passband, M$_G$. Here we see an almost exact parallel to the results using M$_{V_T}$ with a minimum $\Delta$M$_G \sim 0.2$ and a maximum of $\sim 0.91$. Once again, calculating the weighted average, we find an average increase of 0.54 magnitude ($\sim8\sigma$) in the Gaia G band. The similarity between the two bands is not surprising due to the consistency between the apparent G and V$_T$ bands (see Figure \ref{fig:G_Vt_Comp_Clean}). There is clearly a significant brightness increase, exhibited in both photometric systems which we will investigate in more detail in Section \ref{sec:Discussion}.

A similar analysis for the difference in \btvt \ for B and Be stars is shown in the bottom panel of Figure \ref{fig:PhotometryDifferences}. We find a maximum absolute difference of 0.05 mag between B and Be with a weighted average difference of 0.002 mag ($\sim1\,\sigma$). There does not appear to be any significant difference between B and Be stars in the \btvt \ band. However, there is a slight change in trend across the spectral types with Be stars appearing redder than the B stars for earlier spectral types and bluer or similar for later types.

The increased reddening of Be stars at earlier spectral types is more apparent when we compare the differences for \bp . We find a maximum difference of 0.22 mag between B and Be stars which is approximately 4 times greater than we see for \btvt . Overall we find $\Delta$(\bp) indicates an increased reddening across all spectral types compared to $\Delta$(\btvt). The increase in reddening is also apparent from the higher weighted average of -0.03 ($\sim6\sigma$) which is 15 times greater than for $\Delta$(\btvt). Similarly to \btvt, there appears to be a trend of increased reddening for earlier spectral types which lessens for later types.

We conclude this section by stating that the Be stars are indeed brighter than the B stars, while at earlier spectral types, the Be stars also appear to be redder. Additionally, whilst Gaia and Tycho exhibit a similar brightness enhancement, the colour differences  appear to be greater for \bp \ than for \btvt \ . Below we will discuss various mechanisms that could be responsible for these differences.

\section{Discussion}\label{sec:Discussion}

We have used the latest Gaia data release (DR3, \citealt{GaiaDR3}) to study the photometric properties of B and Be stars. In addition, the objects were drawn from the Bright Star Catalogue meaning that our sample thus contains the brightest, closest and least reddened objects. The combination of high precision Gaia-derived distances and minimal uncertainties in the extinction provide the highest quality absolute magnitudes derived for these stars to date. For the B and Be luminosity class V objects, we have presented absolute magnitudes in the Tycho $V_T$ and Gaia $G$ bands, as well as intrinsic \btvt \ and \bp \ colours in Tables \ref{tab:BV_Avg_photometry} and \ref{tab:BVe_Avg_photometry}.

\subsection{Brightness and Colour Excess} \label{sec:Brightness_Colour_Excess}

\textit{Brightness ---} As discussed in the introduction, the brightness enhancement of Be stars compared to their non-emission counterparts has been known for many years and found using various methodologies. For example in \citet{Zorec91} Be stars were found to be brighter by up to 0.8\,mag through comparing the photometric differences of B and Be stars assuming they had the same underlying stellar model. Further studies using Hipparcos such as \citet{Weg1V} and \citet{Zhang} find Be stars to be brighter in the V-band by around 0.5-1 mag yet the average magnitudes reported in the same spectral type can vary significantly (see Table \ref{tab:LitComp}) due to different treatments of extinction and parallax.

\cite{BriiotAbsV} found that the brightness enhancement of Be stars appears to decrease toward earlier spectral types whereas \citet{Zorec91} found the brightness enhancement increases for early-type stars. The results from both of these studies were subject to increased uncertainties in their early-type photometry due to less accurate parallax estimations and/or reduced sample sizes. Fortunately, using our B and Be samples we are in a position to test both hypotheses. Our results seem to indicate a generally consistent brightness enhancement between B1-B9 in both V$_T$ and G, consistent with \citet{Weg2V} and \citet{Zhang}. Therefore, with this sample, we confirm that Be stars are indeed brighter on average by $\sim\,$0.5 magnitudes, consolidating the findings of previous studies and confirming the presence of this enhancement in Gaia photometry. 

Before we discuss the implications of this result, let us first assess the potential impact of the transient nature of the Be phenomenon (see Section \ref{sec:Intro}) on our results. It could be that a ``Be'' star at the time of spectral classification had lost its disc by the time of the Gaia observations, while an object classified as ``B'' may now be surrounded by a disc.  The observed photometric differences would then in fact be a lower limit as the B contaminants would act to bring the Be star values down to that of an uncontaminated B sample while the opposite would be the case for the Be stars within the B sample. We conclude that any phase-changing of the objects under consideration means that the observed photometric differences between Be and B stars are possibly even larger than observed. 
It should be noted that the observed small differences between the $G$ and $V_T$ photometry (a 1$\sigma$ spread of $\sim$0.06 mag, see Section~\ref{sec:SampSelect}) in our sample should act to alleviate any concerns about potential contaminants.

\textit{Colour Excess ---} We additionally investigate the photometric colours of Be stars, as they are both predicted \citep[see e.g.][]{GravityDarkening} and inferred \citep[see e.g.][]{Moujtahid1999,Keller2000,Milone2018, Schootemeijer2022} to be redder than non-Be stars. While many theoretical models provide {\it B~-~V} colours, very few studies actually report on those colours. Reports that Be stars are redder often consider much wider wavelength ranges, extending both to the blue and to the $I$ band and redder. Our results indicate that Be stars are negligibly redder than B stars in Tycho \btvt\,($\sim$0.002 mag) however they are indeed redder than B stars in Gaia \bp\, ($\sim$0.03 mag).  

One explanation for this discrepancy may 
arise from the different wavelength regimes being probed. The Gaia R$_P$ band operates between $\sim$600\,--1000\,nm with an effective wavelength of 783\,nm \citep{EDR3Photometry} whereas the Tycho V$_T$ band encompasses both a smaller wavelength range ($\sim$450\,--680\,nm) and a shorter central wavelength \citep{Bessell2000}. If we are seeing redder colours for the Be stars then the mechanism(s) responsible may emit significantly at wavelengths beyond 680\,nm.

In addition, there appears to be a potential correlation between the observed colour difference and spectral type with Be stars appearing increasingly more red towards early spectral types as seen in Figure \ref{fig:PhotometryDifferences}. This trend is tentatively present in both Tycho and Gaia colours however we find that it is most strongly seen in \bp\,. Therefore, we focus the following discussion on Gaia colours. 

Be stars with earlier spectral types e.g. B1e-B6e (excluding B3e) are typically redder than their non-Be counterparts by $\gtrsim0.05$\,mag. In contrast, late type stars, B8e and B9e, are only redder by $\sim 0.01$\,mag. The colour difference between B and Be stars seems to increase gradually from $\sim 0.01$\,mag for B9e up to $\sim 0.07$\,mag for B6e stars and finally to the largest difference in colour is $\sim 0.22$\,mag for B1e stars. Although this increase is not monotonic, there appears to be a sharp increase between B1 and B6 which begins to plateau at late spectral types (B7-B9).

Several suggestions have been put forward in the literature to explain these photometric effects which we will discuss below.

\subsection{Rotational Velocity and Gravity Darkening}\label{sec:GravDark}

A key property of Be stars is their extreme rotational velocity which is often reported to be close to critical \citep{Rivinius_review}, but we note that after considering all selection biases, \citet{Zorec2016} conclude that this may not necessarily be the case for all Be stars. As illustrated by \citet{GravityDarkening}, very high rotational speeds can have a significant effect on the observed magnitudes and colours through gravitational darkening at the equator. The expected effect is a reddening of the star i.e. a higher {\it B -- V} (up to 0.04 mag in the edge-on case) and a brighter M$_V$  by up to 0.6 mag, when viewed pole-on \citep[see also][]{Collins1991}. These are the maximum values for the extreme cases of both inclination and rotational velocity. For intermediate inclinations, the rapidly rotating stars are computed to be brighter and redder than their non-rotating counterparts, but the excursions in the colour-magnitude diagram will be less large.

\begin{figure}
    \centering
   \includegraphics[width=0.99\linewidth]{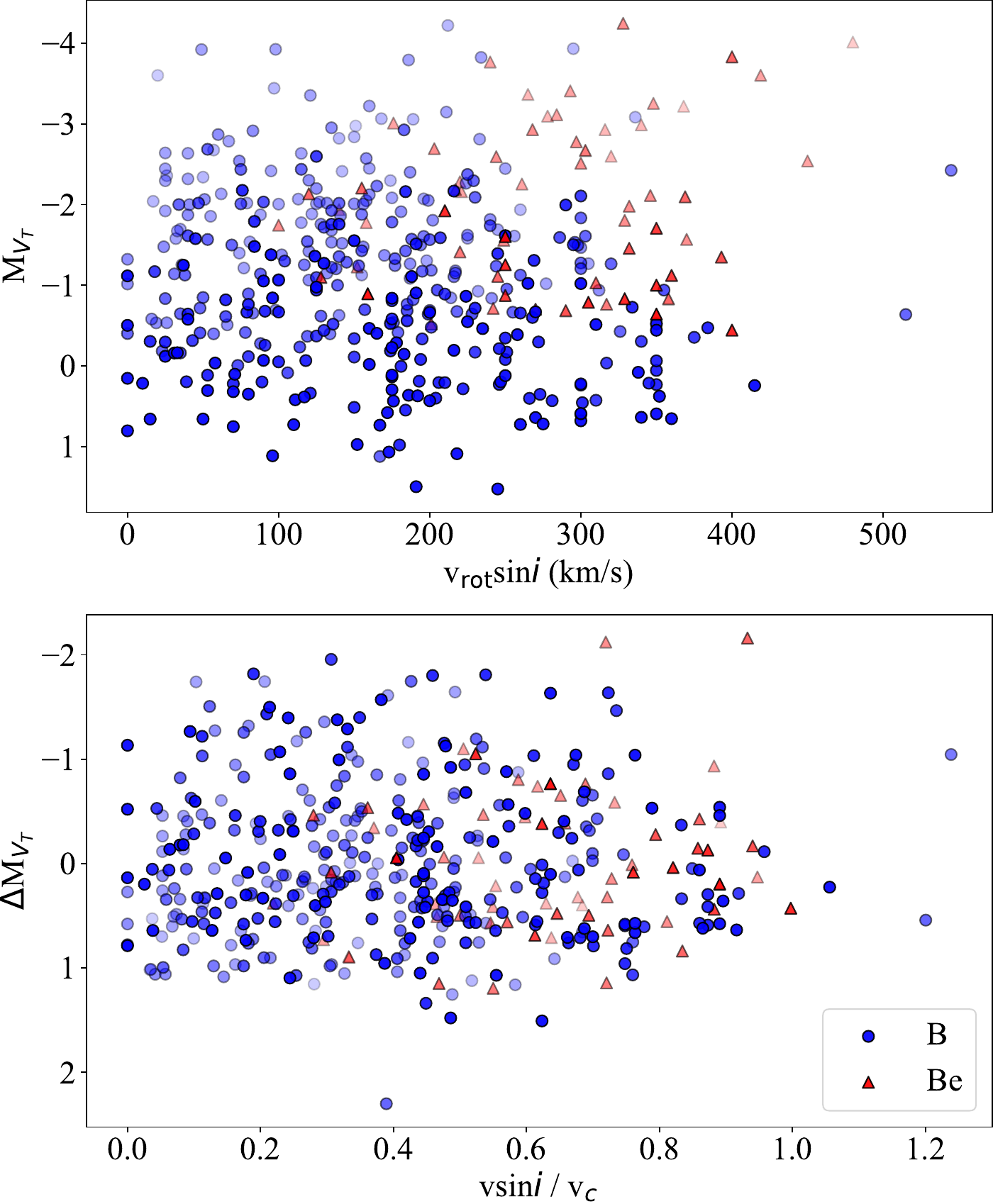}
    \caption{\textbf{\textit{Top:}} The absolute V$_T$ magnitude as function of observed rotational velocity. The B stars are denoted by the filled circles while Be stars are indicated by the triangles with darker colours indicating later spectral types. \textbf{\textit{Bottom:}} The difference between the observed and average absolute V$_T$ magnitude for each spectral type as fraction of the critical velocity for the respective spectral types (taken from \protect\citealt{GravityDarkening}). B and Be stars follow the same pictorial representation as in the top panel. We note that some objects have rotational speeds greater than their assumed critical velocity, this likely reflects uncertainties in either the spectral class or the derived vsin$i$. 
    }
    \label{plotrot}
\end{figure}

With our dataset, we are in the position to investigate the role of gravity darkening for rotating BV and BVe stars empirically. The BSC and its Supplement provide rotational velocities for 55 per cent of our objects, while spot-check comparisons of the reported velocities with those in \citet{BeSS} would indicate uncertainties of order 10-20 per cent, which, as we shall see below, will not affect our main findings.  

The top panel of Figure~\ref{plotrot} shows the absolute $V_T$-band magnitude of both B and Be stars as function of the observed rotational velocity. The later type stars (denoted by darker symbols) are fainter and earlier type objects are brighter, as can be seen in the previous colour-magnitude diagrams. Likewise, and as discussed in Section \ref{sec:B_Be_comp}, the Be stars are typically brighter than the B stars. Figure~\ref{plotrot} also confirms they have larger observed rotational velocities;  hardly any Be objects have projected speeds less than 150 kms$^{-1}$, whereas many B stars do. 

In the bottom panel we compare the difference in absolute magnitude of all stars with the average for their spectral type (See Tables \ref{tab:BV_Avg_photometry} and \ref{tab:BVe_Avg_photometry}) against vsin$i$/v$_c$, where v$_c$ is the critical velocity for each spectral type \citep[see Table 1,][]{GravityDarkening}. The first notable finding is that although Be stars are not found at low velocities, we find both B and Be stars at the higher velocities. In a similar vein, the Be stars are brighter than the B stars, but the range in brightness covered by the Be stars is also covered by B stars. The low number of Be stars at high vsin$i$/v$_c$ appears consistent with Zorec et al.'s 2016 conclusion that Be stars do not rotate at critical velocity; assuming they would all be oriented randomly, the distribution over cos$i$ would be uniform - predicting a much larger number of objects at high vsin$i$/v$_c$ than observed. A striking observation is that there does not seem to be a trend of brightness with rotation velocity for either the full sample, the B and Be stars separately, or for the individual spectral types\footnote{The outliers (2 Be stars are brighter by more than 2 magnitudes than the average and a B star is fainter by 2 magnitudes spring to mind) are probably misclassified in spectral type. It is not uncommon that a small number of objects have been assigned the wrong luminosity class. For example, \citet{Oudmaijer1999} found that 10 per cent of all K0V stars were brighter than expected and more consistent with lower surface gravity luminosity class IV or III stars.}. Previous studies using smaller samples and limited spectral type ranges (e.g. 10's of objects in B2Ve, \citealt{Zhang}; B2-B9, \citealt{Jaschek98}) already alluded to this result, yet here we use not only a larger sample, but thanks to Gaia and the use of extinction corrections independent of spectral type, arguably the best intrinsic magnitudes obtained thus far. Furthermore, we also find no correlation with \btvt \ , \bp \ and M$_G$.

We now investigate the relationship between M$_{V_T}$ and vsin$i$ within a given spectral type more quantitatively. By focusing exclusively on the non-emission B stars, we minimise any effects intrinsic to the nature of Be stars themselves. To this end, we calculate a line of best fit considering all sources within 5$\sigma$ of the average M$_{V_T}$ for a given spectral type. By implementing this selection we aim to mitigate the effects of any possible spectral misclassifications. If no correlation is found, we would expect a gradient of zero to be returned by the fit.

\begin{figure}
    \centering
    \includegraphics[width=\linewidth]{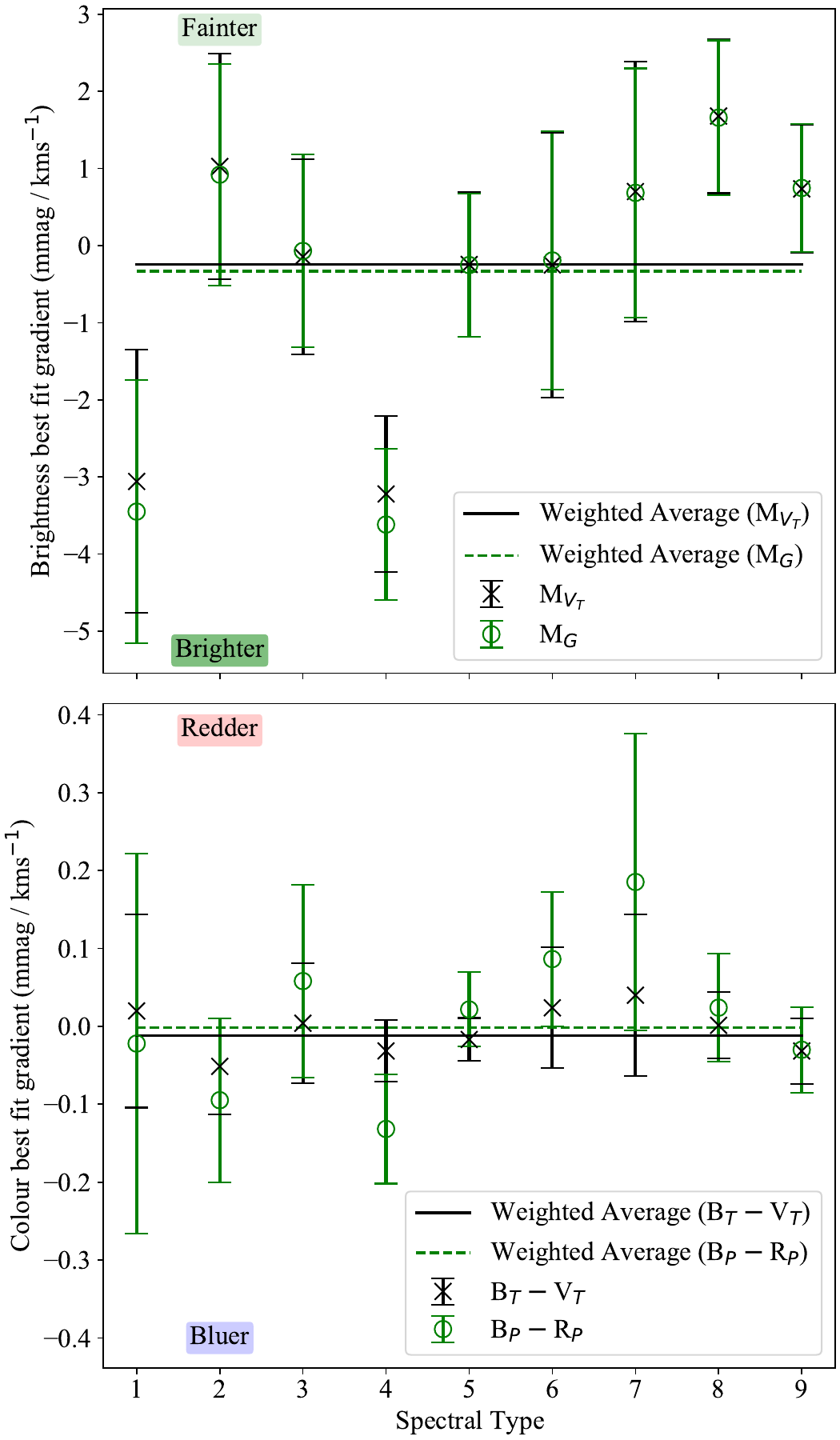}
    \caption{\textbf{\textit{Top:}} The calculated gradients of B stars in each spectral type for the correlation of M$_{V_T}$ and M$_G$ against vsin$i$ shown as black crosses and green open circles, respectively. We indicate the weighted average for M$_{V_T}$ and M$_G$ as the solid and dashed lines respectively. We indicate the regions of increasing and decreasing brightness through labels at the top and bottom of the panel. \textbf{\textit{Bottom:}} The calculated gradients of B stars in each spectral type for the correlation of \btvt \ and \bp \ against vsin$i$ shown as black crosses and green open circles, respectively. We indicate the weighted average for \btvt \ and \bp \ as the solid and dashed lines respectively. We indicate the regions in which stars are considered bluer and redder through labels at the top and bottom of the panel.}
    \label{fig:RotVel_gradients}
\end{figure}

We present the calculated gradients for M$_{V_T}$ and M$_{G}$ against vsin$i$ in the top panel of figure \ref{fig:RotVel_gradients}. For most spectral classes, the slope is consistent with zero, meaning that there is no change in brightness of the objects with rotational velocity.  The  B1 and B4 stars exhibit a different behaviour than other spectral types, where the slope is approximately 2$\sigma$ (B1) and 3 $\sigma$ (B4) from zero. We note that with 15 and 16 objects respectively, these are  the two least populated samples within this analysis. A weighted average considering all spectral classes returns a gradient of -0.2 $\pm$ 0.4 and -0.3 $\pm$ 0.4 mmag/kms$^{-1}$ for M$_V$ and M$_G$ respectively, which indicates no significant trend in brightness enhancement for increasing vsin$i$.

As Be stars are predicted to be redder in some cases \citep{GravityDarkening}, 
we perform the same analysis for the extinction corrected colour as a function of rotational speed, summarised in the bottom panel of Figure \ref{fig:RotVel_gradients}. We find that all spectral classes have a gradient (within errors) consistent with zero indicating that there is no strong effect on the observed \btvt \ nor the observed \bp \ due to gravitational darkening. We find that changes in \btvt \ with vsin$i$ are very small with a maximum absolute gradient of 0.05 mmag/kms$^{-1}$  and a weighted average gradient of -0.01 $\pm$ 0.02 mmag/km$^{-1}$ which is consistent with a slope of zero. Curiously, for \bp \ , we find a maximum absolute gradient of 0.2 mmag/kms$^{-1}$ which is much higher than for \btvt \ yet a significantly lower weighted average of -0.002 $\pm$ 0.3 mmag/kms$^{-1}$. In any case, there is no strong evidence for a correlation between the observed colour and the rotational velocity and thus it appears that gravitational darkening does not have a strong reddening effect on our objects. We note that \citet{GravityDarkening} suggest that the observationally derived rotational velocities for critically rotating objects can underestimate the true rotational velocity. We cannot exclude this from the above exercise, but do point out that we would still expect to see a trend of observed properties with vsin$i$.

As a function of rotational velocity, highly inclined objects are predicted to be redder than the pole-on objects, while the pole-on objects would be brighter than the edge-on ones, yet neither of those trends is observed. Moreover, the Be stars are brighter than even the rapidly rotating B stars, while the brightness difference of 0.5 magnitude would only be reached by the fastest, edge-on, rotators.  This leads us not only to conclude that fast rotation is not responsible for the different locations of  the B and Be stars in the HR diagram, but also provides further support to the finding of \citet{Zorec2016} who posited that Be stars do not rotate close to critical. 

To conclude this exercise - although it is clear that a rapidly rotating star can be brighter and/or redder when rotating critically \citep{GravityDarkening}, we find that not many B and Be stars rotate close enough to the critical velocity to exhibit the effect in such a manner to explain the brightness differences between B and Be stars (cf. \citealt{Zorec2016}). We also demonstrate empirically that the effect is not present in the current sample.

\subsection{Binarity}\label{sec:binarity}

 An additional mechanism we can investigate is whether differences in binarity might explain the difference in the positions in the HR diagram between the Be and B stars. The presence of an unresolved secondary companion is bound to make the visual magnitude of a system brighter, while the colour of the system may be redder if the companion is a lower mass main sequence star. This would qualitatively explain our observations, however the observed magnitude difference between B and Be stars is of order 0.5 magnitudes. As an equal brightness binary would make an object 0.75 magnitude brighter, it is arguably a tall order to fully explain the magnitude  difference. We not only need a substantially different binary fraction for the B and Be stars, the brightness difference between a binary and a non-binary system needs to be as close to the maximum possible. The angular resolution limit of Tycho-2 and Gaia is about 0.8 and 0.6 arcsec respectively (\citealt{Hog2000b}, p. 382; \citealt{Lindegren2021}), so the question is whether we can say anything about possible subarcsecond companions.

Fortunately, Gaia  provides information that allows us to establish whether a Gaia source may be such a binary.  For example, the renormalised unit weight error (RUWE, 
\citealt{Lindegren2021}) is a powerful quality indicator of the Gaia astrometry. It measures the goodness-of-fit for the astrometric solution and is sensitive to the presence of extended emission or binary companions.  Given that B and Be stars are, to all intents and purposes, point sources at the Gaia resolution (the Be star discs are of order less than a milli-arcsecond in size,  \citealt{Quirrenbach1997,Wheelwright2012}), the RUWE parameter can be used as an indication of binarity  \citep{Belokurov2020,Penoyre2020}.  In addition, the so-called Proper Motion Anomaly (PMa, cf.\citealt{Kervella2022}), which measures the deviation of the proper motion measured in a short term (for example during period the DR3 astrometry was taken) from that determined based on a longer-term (such as the time that passed between the period in which Hipparcos determined a star's position and Gaia), is a powerful indicator of the presence of a binary. \citet{Kervella2022} computed the difference between the long-term Hipparcos-Gaia proper motion vector and the short-term Gaia DR3 proper motion vectors and provided a catalogue of PMa measurements. A typical detection would have a signal-to-noise ratio (SNR) larger than 3. \citet{Dodd2024} investigated the ability of both the RUWE and the PMa to detect companions, and empirically found that binaries with separations between 0.02 and 1 arcsecond could be retrieved. With the comparable resolution of Gaia and Tycho this is therefore a useful diagnostic for unseen companions. 

\begin{table}
    \centering
    \begin{tabular}{lcccc}
    \hline
   Criteria & $\Delta$ M$_{V_T}$ & $\Delta$ \btvt & $\Delta$ M$_G$ & $\Delta$ \bp \\
  & mmag & mmag & mmag & mmag \\
\hline
      SNR<3 & 50 $\pm$ 30 &  -2 $\pm$ 2 & 50 $\pm$ 30 &  -3 $\pm$ 3 \\
      SNR>3 &    -80 $\pm$ 40 &     2 $\pm$ 2 &  -80 $\pm$ 40 &     4 $\pm$ 4 \\
      SNR>5 &  -120 $\pm$ 40 &    2 $\pm$ 3 &  -120 $\pm$ 40 &     5 $\pm$ 5 \\
      \hline
 RUWE < 1.4 &    40 $\pm$ 30 &   -1 $\pm$ 2 &   40 $\pm$ 30 &   -2 $\pm$ 3 \\
 RUWE > 1.4 &    -100 $\pm$ 50 &   2 $\pm$ 4 &   -90 $\pm$ 50 & 7 $\pm$ 6\\
\hline
\end{tabular}
    \caption{The average difference of all B and Be integer spectral types from their derived average values (e.g. $\Delta$M$_G$ = $\overline{\text{M}_G}$ - M$_{G, \text{Obs}}$) shown in tables \ref{tab:BV_Avg_photometry} and \ref{tab:BVe_Avg_photometry} respectively, for the given criteria.}
    \label{tab:RUWE_SNR}
\end{table}

{\it The RUWE parameter:} Table \ref{tab:RUWE_SNR} shows the average difference of a given absolute magnitude (e.g. $\Delta$M$_G$) or colour (e.g. $\Delta$\bp \ ) with the average value for each spectral type presented in tables \ref{tab:BV_Avg_photometry} and \ref{tab:BVe_Avg_photometry} for B and Be respectively. As discussed by, for example \citet{Dodd2024}, a value larger than 1.4 is a strong indicator of binarity. We see that for objects with RUWE$\,>\,1.4$, both their M$_{V_T}$ and M$_G$ absolute magnitudes are brighter at the 2$\sigma$ level compared to those with a smaller RUWE value. An increased brightness would be expected from stars with a binary companion compared to those without, however, we find a rather low average increase of $\sim -$0.1 magnitudes compared to the observed 0.5\,mag.  Even with our comparatively large uncertainties we cannot reconcile the brightness increase due to possible binarity with our observed increase from the Be stars. With regards to a possible reddening of the observed colour due to the contribution of the companion, we do not find any evidence in both \btvt \ and \bp ,  any differences are at the 1$\sigma$ level at best.

{\it The Proper Motion Anomaly:}  We use the SNR derived by \citet{Dodd2024} to separate our samples into 'binary' and 'non-binary' using a binary threshold of SNR$\,>\,3$ and a more conservative threshold of SNR$\,>\,5$. The results here echo those found above with the RUWE parameter with a slight brightness increase in the binaries and no significant colour difference. The brightness increase in M$_{V_T}$ and M$_G$ is more significant (at the 3$\sigma$ level) for the more conservative threshold, yet the difference is at the $\sim$0.1 magnitude level.

 As discussed at the beginning of this Section, only in the extreme case that all Be type objects are in an unresolved, equal-brightness binary system, would we be able to explain the magnitude difference between Be and B-type stars. In fact, it would require a 100 per cent binary fraction for the Be stars and a 0 per cent binary fraction for the B stars, neither of which are observed. If any equal-brightness binaries would be at even closer separations they might not be detectable by the Gaia RUWE and PMa approaches. However, there is no evidence that this is the case, it would for example imply that Be stars would have clear evidence for double-lined spectra. The number of spectroscopic binaries (either SB1 or SB2) does not reach these levels however (see for example \citealt{Abt1984} for a search for spectroscopic Be binaries and \citealt{Bodensteiner2020} for a search for Main Sequence companions to Be stars). 

To conclude, we have shown that although the presence of a binary companion may contribute to an increased brightness, binarity alone cannot explain the fact that Be stars are on average 0.5 magnitudes brighter than their B-type counterparts. Additionally, we find that binarity does not achieve, at least in our sample, the colours necessary to explain the redward shift of Be stars compared to B stars seen in \bp \ (Figure \ref{fig:PhotometryDifferences}).

\subsection{Contribution of disc to spectral energy distribution} 

The Be stars' circumstellar discs are well known to emit continuum emission, mostly due to free-free and bound-free emission by hydrogen. The contribution of this emission dominates at the longest wavelengths and is observed to strongly emit at near-infrared and longer wavelengths (e.g. \citealt{Dougherty1994,Waters1987}). Further studies, using non-LTE radiative transfer models such as  \citet{Carciofi2006} estimate the effect of a gaseous circumstellar disc on the resultant spectral energy distribution of a Be star, once again confirming that the contribution from the disc increases at progressively longer wavelengths. It is hard however to directly assess the contribution of the disc to the $B, V$ bands in the spectrum. As the discs are small, we would need to use optical/infrared interferometry to separate the disc from its host star and determine their respective contributions to the total flux of the objects. This has been possible, and been carried out for near-infrared observations (e.g. \citealt{Touhami2013}) and H$\alpha$ observations (\citealt{Quirrenbach1997}), where the disc contribution to the total flux is relatively large. However, observations at the  optical wavelengths have been sparse, and are mostly aimed at binary studies rather than the smaller, fainter and therefore possibly undetectable discs at these shorter wavelengths (e.g. \citealt{hutter2021})

Hence, we have to resort to indirect observational measures of the disc's contribution to the total flux of the Be star system at the  wavelengths where Gaia and Tycho operate. As discussed above, the excess becomes progressively larger at longer wavelengths. It is then comparatively easy  to determine its contribution by comparing the observed spectral energy distribution with that of a naked star. However, this is not necessarily a straightforward exercise at optical wavelengths. 
In contrast, any excess emission is readily observed when considering any variability of the objects, and especially the situations where discs are formed during a burst of mass loss from the star. \citet{Haubois2012} worked out the contribution of a disc that is being built up and found that stars around which a disc develops can become brighter by 0.3 magnitudes and redder by 0.1 magnitude when the disc is close to face-on (see also \citealt{Bernhard2018,Rimulo2018}). Disc build-ups were already inferred from photometric variations with similar changes in brightness and magnitude (e.g. \citealt{dewit2006}) and have now been regularly observed in not only photometry, but polarization and H$\alpha$ emission line strengths as well (\citealt{Jones2013,Marr2021,Rast2024}).

While the contribution of the disc to the total flux can reach 0.3 magnitudes (and sometimes even greater), we should point out that  the average excess found in our sample  is 0.5 magnitudes.  This would be among the most extreme when considering both observations and models of the circumstellar disc emission. It would thus appear that the disc can only be a part of the explanation for the brightness difference between B and Be stars. Regarding the colours, most predictions report modest (smaller than 0.1 magnitude) colour differences, which are not incompatible with those observed here.  For example, if we assume the stars in our sample are randomly oriented, the 0.1\,mag colour difference may be averaged out to values closer to 0.03\,mag as in the $i=70$ degrees model of \citet{Haubois2012}. While this value is based on the Johnson {\it B-V}, there are qualitative similarities to the Gaia colours. Therefore, disc build-up alone cannot reproduce the observed brightness enhancement and thus, the disc may not be solely responsible for the photometric differences of Be stars.

\subsection{Evolution of the central star}

Another process that could explain the fact that Be stars are typically brighter is the evolution of the central star. For example, B stars at the Terminal Age Main Sequence (TAMS) are brighter by one magnitude or more than when at the Zero Age Main Sequence (ZAMS).  Depending on the stars' masses and rotational speeds, they can also be redder than on the ZAMS.  The main requirement for the (mass loss leading to a) disc, is that the star rotates close to its critical velocity. Unless all Be stars were born rapidly rotating (a notion that is not currently favoured \citealt{Rivinius2024}), they would have to be spun-up at some point. For example, in the single star scenario, the spin-up occurs during the main sequence phase (e.g. \citealt{Maeder2000,Ekstrom2008,Mombarg2024}), while in the binary scenario the star gains both mass and angular momentum from interaction with a companion (e.g. \citealt{Klement2024}). Either process will take time, and it could be expected that Be stars would be found  - on average - further in their Main Sequence evolution than non-Be stars (e.g. \citealt{Sibony2024}).

Here it may be interesting to note that \citet{Zorec2005} derived fundamental parameters spectroscopically (independent of the disc emission, \citealt{Fremat2005}) for a sample of Be stars. They found that whereas the earliest Be-type stars were located close to the ZAMS, the later type objects were more evolved up the Main Sequence. We can thus immediately infer that the later type Be stars can be expected to be brighter than their B-type counterparts which will be distributed more uniformly over the Main Sequence. We  conclude, therefore, that the  brightness of these objects is due to the stars' evolved nature and the colour may be explained by the excess disc emission. 

This still leaves the finding that the early type Be stars are also an average of 0.5 magnitude brighter than the non-Be stars, but are perhaps not as evolved on the Main Sequence. Inspection of Figure~\ref{fig:StilHRD_B_BE} reveals that the spread in excesses is larger for the early type stars; they may be closer to the ZAMS (as found by \citealt{Zorec2005}), but when there is an excess, the excess emission is much larger. This may not come as a surprise as the H$\alpha$ emission and near-infrared excesses are much stronger for early-type Be stars, indicating they have much larger and denser discs (e.g. \citealt{Banerjee2021,Dougherty1994}). We should note that the observations cited earlier do not often report larger than, say, 0.5 magnitude variations, and therefore would rule out that the discs contribute substantially to the observed excess emission. A counter argument could be that the early-type objects constitute a minority of Be stars and that the average observed for the entire class may not be representative for the early B-types. To this we add that these objects do seem to be redder than their B-type counterparts (as per Figure~\ref{fig:PhotometryDifferences}), in contrast to the cooler objects discussed above. Therefore, we can reach a similar conclusion for the early-type Be stars as for the later type Be stars; both the emission from the disc and the evolved star contribute to the excess emission, with the proviso that  the disc is likely to dominate this excess for the hottest objects.

\section{Summary and Conclusions}\label{sec:ConcludingRemarks}

In this study we have made use of Gaia and Tycho which offer the most precise astrometric and photometric data available, enabling us to mitigate the effects of extinction and distance dependent errors for our comparative photometric study of B and Be stars. We have used the BSC to further minimise the uncertainties in our derived photometric properties allowing us to accurately study the differences between classical Be stars and their non-emission counterparts. We found that the Be stars are brighter by about 0.5 magnitudes than the B stars for every spectral type. We also found that the earliest type Be stars are redder than the B stars.

In order to understand why these photometric differences arise, we considered the effect of several different mechanisms on our sample. We investigated the effect of gravity darkening both qualitatively and quantitatively and found it cannot reproduce the observed brightness enhancement and colours of Be stars, at least for our sample. We then discussed the potential effects of binarity and found that an unseen companion can, theoretically, cause the observed brightness enhancement of 0.5\,mag if the companion is of equal mass, but this is not observed when studying the potential binary systems in our sample. Moreover, binarity alone cannot explain the observed colour difference we see in \bp\, and \btvt.

When considering the contribution of the disc, we find that its emission would appear to not be sufficient on its own to explain the large brightness excess of, on average, 0.5 magnitudes for most objects. Therefore, we find that the combination of both disc emission and stellar evolution along the Main Sequence best explains our observations. 

Finally, we can reconcile the increased colour excess of Gaia \bp\, compared to Tycho \btvt\, due to the fact that the R$_P$ band probes longer wavelengths in which the disc also emits significantly (see Section \ref{sec:Brightness_Colour_Excess}). Therefore, any observational effects caused by the disc may be more easily observed in \bp\, due to its greater coverage of potential disc emission.

\bigskip

We summarise the main findings of our work as follows:

\begin{itemize}
\item[--] We present the first Gaia colour magnitude diagram comparing BV and BVe stars and find that Be stars occupy a region  above (brighter) and marginally redward of the B-type main sequence.

\item[]

\item[--] We provide, for the first time, average photometric properties for both B and Be spectral types in Gaia M$_G$ and \bp. We also present intrinsic Tycho M$_{V_T}$ and \btvt \ in Tables \ref{tab:BV_Avg_photometry} and \ref{tab:BVe_Avg_photometry}.

\item[]

\item[--] We find that Be stars have a typical brightness enhancement of $\sim$0.5 magnitudes in both M$_G$ and M$_{V_T}$ consolidating that this effect is present in multiple photometric systems.

\item[]

\item[--] We present the first evidence that Be stars appear to be slightly redder by on average 0.03 magnitudes in the Gaia \bp \ band, with the earliest spectral types showing the strongest effect. However, this reddening is much less prevalent in Tycho \btvt \ .

\item[]

\item[--]The above  indicates that whatever mechanism is responsible for the reddening, it is more prominent in the \bp \ range than it is in \btvt . 

\item[]

\item[--] We find no evidence that the effects of gravitational darkening due to fast rotation nor binarity can explain the observed photometric properties of our sample of Be stars. It would appear that the brightness enhancement is due to the combination of the stars becoming brighter when evolving up the Main Sequence and the contribution of a circumstellar disc, which in turn explains the increased reddening seen in \bp \  and not in \btvt. 
    
\end{itemize}

\noindent\textit{Final comments} 

\noindent The work as presented here concerns the largest study on the photometric properties of B and Be stars, yet still suffers from the fact that at optical wavelengths we have to infer the emission from disc and star indirectly. In order to accurately understand the contribution of the disc on a star's photometry we require spatially resolved spectro-photometry to allow us to decompose the observed emission into its relative wavelength contributions. To this we add that by analysing individual stellar spectra we can homogenise the spectral type classifications. Indeed, stellar spectra computed for various rotational velocities such as being performed by \citet{Rubio2023} and \citet{Montesinos2024} will allow us to assess how spectral types derived from the observed spectra will be affected by the rapid rotation of the stars.

\section * {Acknowledgments}

IR acknowledges a studentship funded by the Science and Technology Facilities Council of the United Kingdom (STFC).

This work has made use of data from the European Space Agency (ESA) mission
{\it Gaia} (\url{https://www.cosmos.esa.int/gaia}), processed by the {\it Gaia} Data Processing and Analysis Consortium (DPAC, \url{https://www.cosmos.esa.int/web/gaia/dpac/consortium}). Funding for the DPAC has been provided by national institutions, in particular the institutions participating in the {\it Gaia} Multilateral Agreement.
This research has made use of the SIMBAD database, operated at CDS, Strasbourg, France.

\section * {Data Availability}
The data underlying this article are available in the article and in its online supplementary material.

    \bibliographystyle{mnras}
    \bibliography{00_Bibliography} 
    \appendix

\section{Average Photometry}
We present the average values for BV and BVe sources in both Tycho and Gaia photometric passbands in table \ref{tab:BV_Avg_photometry} and \ref{tab:BVe_Avg_photometry} respectively.
\begin{table*}
    \centering
    \begin{tabular}{lccccc}
\toprule
Spectral Type & M$_{V_T}$ & B$_T -$ V$_T$ & M$_{G}$ & B$_P-$ R $_P$ & \# Objects \\
\midrule
    B0V+B0.5V &      -3.3 $\pm$ 0.3 &    -0.23 $\pm$ 0.04 &     -3.1 $\pm$ 0.3 &    -0.21 $\pm$ 0.07 &        13 \\
          B0V &      -4.13 $\pm$ 0.83 &    -0.34 $\pm$ 0.06 &    -3.9 $\pm$ 0.7 &    -0.3 $\pm$ 0.1 &         4 \\
        B0.5V &    -2.9 $\pm$ 0.3 &    -0.18 $\pm$ 0.04 &    -2.8 $\pm$ 0.3 &     -0.15 $\pm$ 0.09 &         9 \\
    B1V+B1.5V &    -2.9 $\pm$ 0.1 &    -0.26 $\pm$ 0.01 &    -2.8 $\pm$ 0.1 &    -0.31 $\pm$ 0.01 &        48 \\
          B1V &     -3.06 $\pm$ 0.16 &     -0.25 $\pm$ 0.01 &     -3.0 $\pm$ 0.2 &     -0.30 $\pm$ 0.02 &        28 \\
        B1.5V &      -2.6 $\pm$ 0.2 &     -0.26 $\pm$ 0.01 &   -2.6 $\pm$ 0.2 &   -0.33 $\pm$ 0.01 &        20 \\
    B2V+B2.5V &    -2.09 $\pm$ 0.07 &   -0.216 $\pm$ 0.004 &    -2.03 $\pm$ 0.07 &    -0.264 $\pm$ 0.007 &       105 \\
          B2V &      -2.18 $\pm$ 0.09 &  -0.219 $\pm$ 0.005 &    -2.11 $\pm$ 0.09 &    -0.271 $\pm$ 0.009 &        68 \\
        B2.5V &      -1.93 $\pm$ 0.01 &   -0.210 $\pm$ 0.004 &   -1.9 $\pm$ 0.1 &    -0.25 $\pm$ 0.01 &        37 \\
          B3V &    -1.59 $\pm$ 0.06 &    -0.192 $\pm$ 0.004 &   -1.55 $\pm$ 0.06 &   -0.236 $\pm$ 0.006 &        97 \\
          B4V &      -1.38 $\pm$ 0.11 &   -0.178 $\pm$ 0.003 &   -1.3 $\pm$ 0.1 &  -0.220 $\pm$ 0.005 &        32 \\
          B5V &    -1.18 $\pm$ 0.06 &   -0.164 $\pm$ 0.005 &   -1.13 $\pm$ 0.06 &    -0.193 $\pm$ 0.009 &       111 \\
          B6V &     -0.99 $\pm$ 0.11 &    -0.141 $\pm$ 0.008 &   -0.9 $\pm$ 0.1 &   -0.16 $\pm$ 0.01 &        40 \\
          B7V &    -0.64 $\pm$ 0.11 &   -0.124 $\pm$ 0.006 &   -0.6 $\pm$ 0.1 &    -0.14 $\pm$ 0.01 &        46 \\
    B8V+B8.5V &   -0.33 $\pm$ 0.07 &  -0.109 $\pm$ 0.003 &   -0.30 $\pm$ 0.06 &    -0.116 $\pm$ 0.004 &       147 \\
          B8V &    -0.36 $\pm$ 0.07 &  -0.109 $\pm$ 0.003 &  -0.32 $\pm$ 0.07 &     -0.116 $\pm$ 0.005 &       141 \\
        B8.5V &     0.3 $\pm$ 0.2 &    -0.10 $\pm$ 0.02 &    0.3 $\pm$ 0.2 &    -0.11 $\pm$ 0.02 &         6 \\
    B9V+B9.5V &   0.02 $\pm$ 0.05 &    -0.066 $\pm$ 0.002 &  0.04 $\pm$ 0.05 &   -0.062 $\pm$ 0.003 &       258 \\
          B9V &  0.02 $\pm$ 0.06 &  -0.075 $\pm$ 0.003 &  0.04 $\pm$ 0.06 &   -0.075 $\pm$ 0.004 &       167 \\
        B9.5V &   0.01 $\pm$ 0.08 &  -0.050 $\pm$ 0.003 &   0.03 $\pm$ 0.08 &   -0.038 $\pm$ 0.005 &        91 \\
\bottomrule
\end{tabular}
    \caption{Average photometry for each non-emission B spectral type, corrected for extinction as discussed in section \ref{sec:Int_Extinction}.}
    \label{tab:BV_Avg_photometry}
\end{table*}

\begin{table*}
    \centering
    \begin{tabular}{lccccc}
    \toprule
Spectral Type & M$_{V_T}$ & B$_T -$ V$_T$ & M$_{G}$ & B$_P-$ R $_P$ & \# Objects \\
\midrule
  B0Ve+B0.5Ve & -3.9 $\pm$ 0.4 & -0.20 $\pm$ 0.05 & -3.8 $\pm$ 0.3 & -0.1 $\pm$ 0.1 & 3 \\
    B0Ve &    -3.62 $\pm$ 0.40 &   -0.15 $\pm$ 0.03 &   -3.6 $\pm$ 0.3 &   -0.01 $\pm$ 0.01& 2 \\
       B0.5Ve & -4.4 & -0.3 & -4.3 & -0.4 & 1 \\
  B1Ve+B1.5Ve &  -3.3 $\pm$ 0.2 &  -0.21 $\pm$ 0.02 &    -3.4 $\pm$ 0.1 &    -0.10 $\pm$ 0.04 &        10 \\
         B1Ve &   -3.31 $\pm$ 0.17 &   -0.21 $\pm$ 0.02 &   -3.4 $\pm$ 0.2 &   -0.08 $\pm$ 0.04 &         9 \\
       B1.5Ve & -3.4 & -0.2 & -3.3 & -0.3 &         1 \\
  B2Ve+B2.5Ve &     -2.5 $\pm$ 0.1 &    -0.19 $\pm$ 0.01 &  -2.4 $\pm$ 0.1 &    -0.19 $\pm$ 0.02 &        29 \\
         B2Ve &     -2.67 $\pm$ 0.16 &  -0.19 $\pm$ 0.01 &   -2.6 $\pm$ 0.2 &   -0.18 $\pm$ 0.03 &        18 \\
       B2.5Ve &    -2.1 $\pm$ 0.2 &  -0.20 $\pm$ 0.01 &    -2.1 $\pm$ 0.2 &   -0.20 $\pm$ 0.03 &        11 \\
         B3Ve &  -2.12 $\pm$ 0.18 &    -0.21 $\pm$ 0.01 &    -2.1 $\pm$ 0.2 &    -0.23 $\pm$ 0.02 &        19 \\
         B4Ve &  -1.91 $\pm$ 0.25 &  -0.16 $\pm$ 0.01 &  -1.9 $\pm$ 0.2 &   -0.15 $\pm$ 0.03 &        11 \\
  B5Ve+B5.5Ve &   -1.8 $\pm$ 0.3 &  -0.17 $\pm$ 0.01 &   -1.7 $\pm$ 0.3 &     -0.16 $\pm$ 0.02 &        13 \\
         B5Ve &  -1.67 $\pm$ 0.26 &  -0.16 $\pm$ 0.01 &  -1.6 $\pm$ 0.3 &    -0.15 $\pm$ 0.02 &        12 \\
       B5.5Ve & -2.7 & -0.22 & -2.7 & -0.26 &         1 \\
         B6Ve &   -1.18 $\pm$ 0.26 &   -0.13 $\pm$ 0.03 &    -1.1 $\pm$ 0.2 &    -0.09 $\pm$ 0.07 &         9 \\
         B7Ve &   -1.56 $\pm$ 0.37 &   -0.15 $\pm$ 0.02 &   -1.5 $\pm$ 0.4 &    -0.15 $\pm$ 0.02 &         4 \\
  B8Ve+B8.5Ve &  -0.8 $\pm$ 0.2 &   -0.13 $\pm$ 0.01 &  -0.7 $\pm$ 0.2 &   -0.11 $\pm$ 0.01 &        11 \\
         B8Ve &   -0.87 $\pm$ 0.18 &  -0.13 $\pm$ 0.01 &  -0.8 $\pm$ 0.2 &   -0.11 $\pm$ 0.01 &        10 \\
       B8.5Ve &  0.3 &  -0.12 &  0.4  & -0.13  &         1 \\
  B9Ve+B9.5Ve &   -0.7 $\pm$ 0.2 &  -0.07 $\pm$ 0.01 &   -0.7 $\pm$ 0.2 &  -0.05 $\pm$ 0.02 &         9 \\
         B9Ve &  -0.84 $\pm$ 0.16 &  -0.08 $\pm$ 0.01 &   -0.8 $\pm$ 0.2 &    -0.07 $\pm$ 0.01 &         7 \\
       B9.5Ve &   -0.35 $\pm$ 0.39 &  -0.02 $\pm$ 0.03 &   -0.3 $\pm$ 0.4 &     0.02 $\pm$ 0.05 &         2 \\
\bottomrule
\end{tabular}
    \caption{Average photometry for each Be spectral type, corrected for extinction as discussed in section \ref{sec:Int_Extinction}.}
    \label{tab:BVe_Avg_photometry}
\end{table*}

\newpage
    \bsp
    \label{lastpage}
\end{document}